# An Analysis of Nonlinear Thickness-shear Vibrations of Quartz Crystal Plates by Two-Dimensional Finite Element Method

Ji Wang[1, *], Yangyang Chen[1], Rongxing Wu[1], Lihong Wang[2], Huimin Jing[1], Jianke Du[1], Yuantai Hu[3], Guoqing Li[3]

[1]Piezoelectric Device Laboratory, School of Mechanical Engineering & Mechanics, Ningbo University, 818 Fenghua Road, Ningbo, Zhejiang 315211, China

[2]Department of Mathematics, School of Sciences, Ningbo University, 818 Fenghua Road, Ningbo, Zhejiang 315211, China

[3]School of Civil Engineering & Mechanics, Huazhong University of Science & Technology, 1037 Luoyu Road, Wuhan, Hubei 430074, China

[*]Corresponding author, e-mail: wangji@nbu.edu.cn

## Abstract

A nonlinear analysis of high-frequency thickness-shear vibrations of AT-cut quartz crystal plates is presented with the two-dimensional finite element method. We expanded both kinematic and constitutive nonlinear Mindlin plate equations and then truncated them to the first-order equations as an approximation, which is used later for the formulation of nonlinear finite element analysis with all zeroth- and first-order displacements and electric potentials. The matrix equation of motion is established with the first-order harmonic approximation and the generalized nonlinear eigensystem is solved by a direct iterative procedure. A backbone curve and corresponding mode shapes are obtained and analyzed. The nonlinear finite element program is developed based on earlier linear edition and can be utilized to predict nonlinear characteristics of miniaturized quartz crystal resonators in the design process.

## 1. INTRODUCTION

Quartz crystal plates in thickness-shear (TSh) vibrations are widely used in resonators for frequency control applications, in which resonant mechanical vibrations are excited by external driving voltages through piezoelectricity. Strong demands for smaller and higher frequency control devices, nonetheless, present challenges in their stability and precision features under complication factors such as thermal and stress fields, for example. When resonators are operated at higher driving voltages, or under various acceleration and temperature fluctuations, it becomes more difficult to maintain their frequency-shifts in a few parts per million (ppm) level around desired frequencies specified in application requirements. Generally speaking, intrinsic nonlinearities are considered to be responsible for these particular frequency variations. Obviously, characteristics of nonlinear vibrations of quartz crystal plates become core elements needed to predict and estimate performance properties of highly stable resonators.

Nonlinear vibrations of plates can be either steady or chaotic through harmonic external forces, which are driving voltages in the case of piezoelectric resonators in circuit applications at radio-frequency (RF). Among those vibration states, periodic and steady nonlinear vibrations are of primary interest due to its existence in the functioning process of device structures. The amplitude-frequency relations, also known as backbone curves, can be obtained by solving undamped systems of linear piezoelectricity for design purpose (Patel, 2008). In resonator applications, the backbone curve is represented by frequency shift under different levels of driving voltage. Such behavior is known as the drive level dependency (DLD) of quartz crystal resonators, which is generally believed to be caused by material nonlinearity and strong electric fields (Patel et al., 2009). Another interesting feature of nonlinear vibrations of quartz crystal resonators involves in internal resonance (Lau et al. 1984; Ribeiro and Petyt, 1999c, 2000), which can lead to the phenomena called “activity dip” in resonators (Patel, 2008). It is also known that the intermodulation could be triggered by a relatively high drive power composed of two neighboring frequencies. In this multimodal state, energy is interchanged between coupling modes. This could be rigorously distinguished from linear couplings of modes, in which uniqueness and superposition of solutions are always valid. The internal resonance usually complicates backbone curves and is therefore more difficult to analyze with existing theory and methods.

The two-dimensional Mindlin plate theory has been one of the mostly used classical methods in the analysis of quartz crystal resonators (Mindlin, 2006; Wang and Yang, 2000), as it was intended for from the beginning. In more than half of a century, its linear theory has served the resonator industry well with solutions to many critical problems in design improvement and product conception. Based on this popular theory for particular quartz crystal resonators required in radio communications and required oscillator circuits, we have developed finite element software with the linear Mindlin plate theory (Wang et al., 1999, 2000), and subsequently improved it recently with some high performance computational features (Wang et al., 2011a). However, as some complicated problems mentioned above have arisen in resonator design and applications, the Minlin plate theory and its finite element implementation must be expanded to include nonlinear analysis for possible design improvements in the reduction of negative effects associated with nonlinear behaviors. As the first attempt, Wang et al. (2009) and Wu et al. (2012) have derived the nonlinear equations for TSh vibrations of isotropic plates in the framework of the Mindlin plate equations, and then solved those truncated plate equations by Galerkin approximation and Homotopy Analysis Method (HAM). It is concluded that the nonlinear theory, although at its aggressive approximation, can provide some explanations about the frequency dependency on the material nonlinearity through couplings of modes at dominant TSh vibrations. Subsequently, they also established nonlinear Mindlin plate theory to study TSh and flexural vibrations of finite AT-cut quartz crystal plates and obtained their backbone curves (Wang et al., 2011b; Wu et al., 2012a, 2012b), again after simplified the coupled nonlinear equations to the sole TSh mode and utilized the Galerkin and HAM. Additionally, TSh vibrations under a strong electric field were also studied with the nonlinear equations of the TSh mode and have formulated and observed relatively strong frequency shift with the presence of electric field (Yang, 1999; Wu et al., 2011). With a similar objective, Patel et al. (2009) investigated the DLD phenomena by the three-dimensional finite element analysis with COMSOL and then followed by experiments from actual products for verification. There is no doubt that the finite element method is an accurate and efficient numerical tool for the nonlinear vibration analysis, which eventually result in a nonlinear eigenvalue system after applying some essential assumptions to obtain the matrix amplitude equation. In the earlier nonlinear finite element analysis of vibrations of beams and plates, Mei (1972, 1973) and Rao et al. (1976) presented some backbone curves of flexural vibrations with simplified assumptions, which are now accepted as standard procedures in the nonlinear finite element formulation and implementation. In addition, some complex techniques involved in nonlinear approximations and iterations for total and updated Lagrangian formulations were adopted in finite element analyses for plate vibrations (Bathe and Bolourchi, 1980; Putcha and Reddy, 1986). Ritz, Galerkin, and harmonic balance methods (HBM) in finite element implementation are widely adopted to eliminate variables in time domain. The equivalency of these three methods has been examined by Lewandowski (1997a). The HBM assumes an expansion of time domain solution into Fourier series, with which the total degree of freedoms (DOFs) will skyrocket by orders of numbers of harmonics stored. To carry out the computation at a better performance, Leung and Fung (1989) discussed the lowest harmonic orders which would be needed to ensure the required accuracy. They suggested that those spatial nonlinear systems need to be linearized and solved iteratively. Lewandowski (1994, 1997b) presented the continuation method, which is also called arc-length method, to study backbone curves of geometrically nonlinear structures. With continuation method again, Ribeiro and Petyt (1999a, 1999b) investigated the amplitude-frequency response and nonlinear mode shapes of isotropic beams and plates. They also paid special attention on the internal resonance and argued that continuation method is an efficient tool in tracing complex backbone curves. Chueng and Lau (1982), Lau et al. (1983), and Chen et al. (2001) analyzed nonlinear structures with incremental harmonic balance method (IHBM), which is a combination of HBM and the incremental Newton–Raphson procedure. Based on a directly iterative linearization method, Han and Petyt (1997a, 1997b) analyzed fundamental flexural vibrations of geometrically nonlinear plates and higher-order modes of laminated plates. In addition, nonlinear vibrations of laminated skew plates were also studied by Singha and Daripa (2007) with the same direct iteration technique.

In present paper, both kinematic and material nonlinearities are incorporated into the nonlinear Mindlin plate equations through the weak form of kinematic and constitutive relations, which were then truncated into the first-order approximation for the formulation of finite element analysis. With the approximation in the first-order harmonic time domain, the resulted matrix amplitude equation is subsequently solved by direct iterations based on given convergent criterion. As a result, one backbone curve of TSh vibrations of an AT-cut quartz crystal plate is obtained from the two-dimensional nonlinear finite element analysis. We also compared mode shapes and frequencies of TSh vibrations from different amplitudes, and examined the distribution of TSh vibrations in plates. This method and software will be considered as an upgraded numerical tool for the analysis and design of quartz crystal resonators with unique capability to handle the DLD and other nonlinear behaviors with the widely used Mindlin plate theory. Apparently, it is a natural growth of our earlier finite element software which was started with the linear theory.

## 2. THE NONLINEAR MINDLIN PLATE EQUATIONS

For an analysis of quartz crystal resonator, the commonly accepted structural model is usually associated with a rectangular quartz crystal plate with coordinate system shown in Fig. 1.

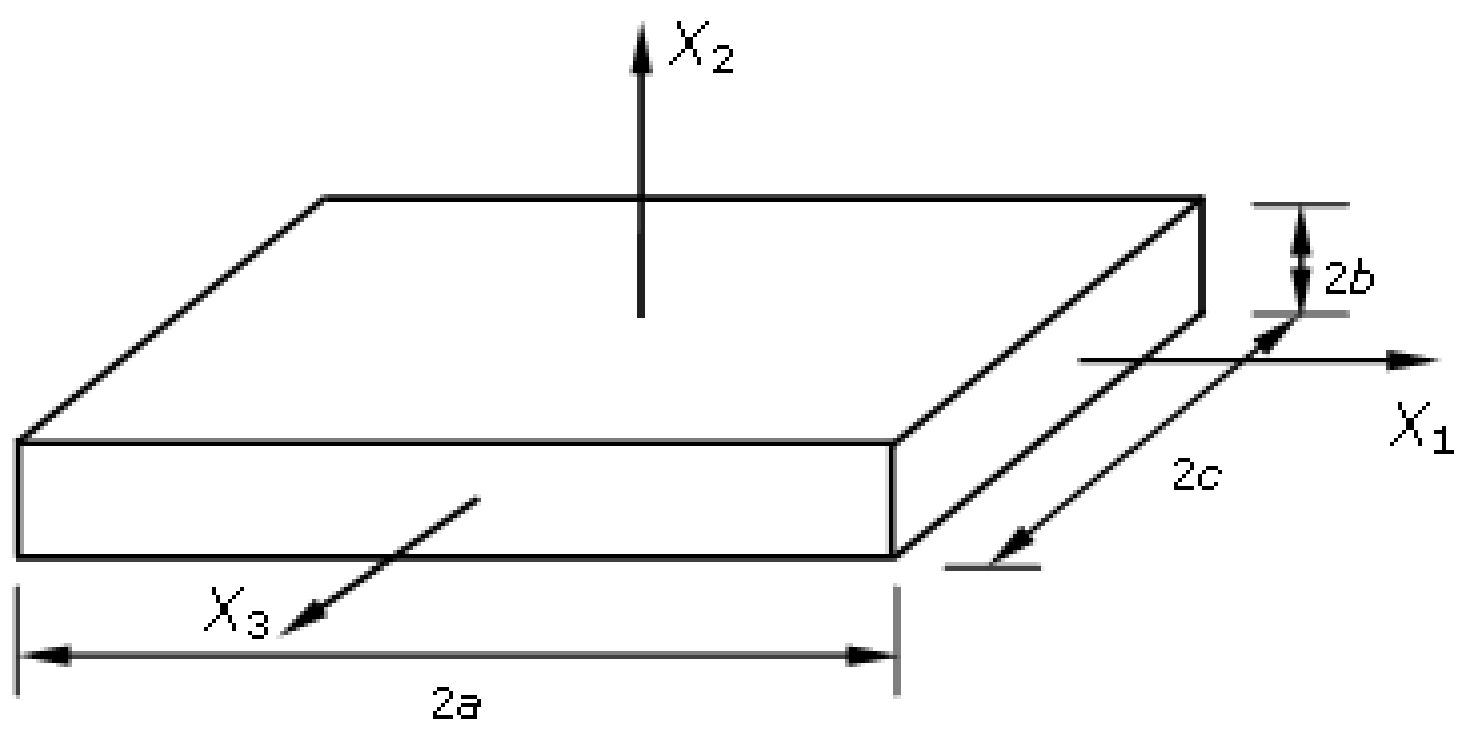

Fig. 1 A rectangular quartz crystal plate with coordinates

Both linear and nonlinear Mindlin plate theory for high-frequency vibrations of piezoelectric plates can be derived with the assumption of expanding all mechanical displacements into power series of the thickness coordinate as (Mindlin, 2006; Wang et al., 2009b; Wu et al., 2012b)

$$u_i(x_1,x_2,x_3,t)=\sum_{n=0}^{\infty}x_2^n u_i^{(n)}(x_1,x_3,t),\ i=1,\ 2,\ 3, \tag{1}$$

Where $u_i^{(n)}$, $x_i$, and $t$ are the $n$th-order displacements, coordinates, and time, respectively. The fundamental TSh vibrations, which is our focus for applications in resonators, is represented by $u_1^{(1)}$.

The $n$th-order two-dimensional Green-Lagrangian strain tensor is defined by (Wang et al., 2009b; Wu et al., 2012b)

$$\begin{aligned} S_{kl}^{(n)} &= \frac{1}{2}\left[u_{k,l}^{(n)}+u_{l,k}^{(n)}+(n+1)\left(\delta_{2l}u_k^{(n+1)}+\delta_{2k}u_l^{(n+1)}\right)\right] \\ &+\frac{1}{2}\left\{\sum_{s=0}^{n}\left[u_{m,l}^{(s)}+\delta_{2l}(s+1)u_m^{(s+1)}\right]\left[u_{m,k}^{(n-s)}+\delta_{2k}(s+1)u_m^{(n-s+1)}\right]\right\},\quad k,l,m=1,2,3, \end{aligned} \tag{2}$$

in which $\delta_{2k}$ is the Kronecker delta. Mathematically, $S_{kl}^{(n)}$ can also be divided into linear and nonlinear parts, as appeared in the first and second lines of (2). It needs to be noted that the linear part of the nonlinear strain here is identical with the strain definition in the linear Mindlin plate theory (Mindlin, 2006). For the implementation of finite element method, the zeroth- and first-order strains could be expressed only with zeroth- and first-order displacements as

$$\begin{aligned} S_{kl}^{(0)} &= \frac{1}{2}(u_{k,l}^{(0)}+u_{l,k}^{(0)}+\delta_{2l}u_k^{(1)}+\delta_{2k}u_l^{(1)})+\frac{1}{2}(u_{m,l}^{(0)}+\delta_{2l}u_m^{(1)})(u_{m,k}^{(0)}+\delta_{2k}u_m^{(1)}), \\ S_{kl}^{(1)} &= \frac{1}{2}(u_{k,l}^{(1)}+u_{l,k}^{(1)})+\frac{1}{2}[u_{m,k}^{(1)}(u_{m,l}^{(0)}+\delta_{2l}u_m^{(1)})+u_{m,l}^{(1)}(u_{m,k}^{(0)}+2\delta_{2k}u_m^{(1)})]. \end{aligned} \tag{3}$$

Their differentiations then become

$$
\begin{aligned}
\partial S_{kl}^{(0)} &= \frac{1}{2}\left(\partial u_{k,l}^{(0)} + \partial u_{l,k}^{(0)} + \delta_{2l}\partial u_k^{(1)} + \delta_{2k}\partial u_l^{(1)}\right) + \frac{1}{2}(u_{m,l}^{(0)} + \delta_{2l}u_m^{(1)})(\partial u_{m,k}^{(0)} + \delta_{2k}\partial u_m^{(1)}) \\
&\quad + \frac{1}{2}(\partial u_{m,l}^{(0)} + \delta_{2l}\partial u_m^{(1)})(u_{m,k}^{(0)} + \delta_{2k}u_m^{(1)}), \\
\partial S_{kl}^{(1)} &= \frac{1}{2}(\partial u_{k,l}^{(1)} + \partial u_{l,k}^{(1)}) + \frac{1}{2}[u_{m,k}^{(1)}(\partial u_{m,l}^{(0)} + \delta_{2l}\partial u_m^{(1)}) + \partial u_{m,k}^{(1)}(u_{m,l}^{(0)} + \delta_{2l}u_m^{(1)}) \\
&\quad + u_{m,l}^{(1)}(\partial u_{m,k}^{(0)} + 2\delta_{2k}\partial u_m^{(1)}) + \partial u_{m,l}^{(1)}(u_{m,k}^{(0)} + 2\delta_{2k}u_m^{(1)})].
\end{aligned} \tag{4}
$$

The differential operator $\partial$ here is symbolic and further operations will be carried out with specific definition through a proper subscript.

From Cauchy stress tensor, the two-dimensional nonlinear constitutive relations are defined by a standard procedure as (Wang et al., 2009b; Wu et al., 2012b)

$$
T_{ij}^{(n)} = \sum_{m=0}^{\infty} B_{mn}\left(c_{ijkl}S_{kl}^{(m)} + \frac{1}{2}c_{ijklst}S_{klst}^{(m)}\right), \quad i,j,k,l,s,t = 1,2,3, \tag{5}
$$

where $T_{ij}^{(n)}$, $c_{ijkl}$ and $c_{ijklst}$ are the $n$th-order stress, the second-order, and third-order elastic constants, respectively, and

$$
\begin{aligned}
B_{mn} &= \begin{cases} 2b^{m+n+1}(m+n+1)^{-1}, & m+n = \text{even}, \\ 0, & m+n = \text{odd}, \end{cases} \\
S_{klst}^{(n)} &= \sum_{p=0}^{n} S_{kl}^{(n-p)}S_{st}^{(p)}.
\end{aligned} \tag{6}
$$

We may also want to separate (5) into linear and nonlinear parts mathematically with the finding that (5) would be the same with stresses appeared in the linear theory by suppressing its nonlinear terms. Through expanding (5), the zeroth- and first-order constitutive relations can be approximated as

$$
\begin{aligned}
T_{ij}^{(0)} &= 2bc_{ijkl}S_{kl}^{(0)} + bc_{ijklst}S_{klst}^{(0)}, \\
T_{ij}^{(1)} &= \frac{2b^3}{3}c_{ijkl}S_{kl}^{(1)} + \frac{b^3}{3}c_{ijklst}S_{klst}^{(1)},
\end{aligned} \tag{7}
$$

where

$$
\begin{aligned}
S_{klst}^{(0)} &= S_{kl}^{(0)}S_{st}^{(0)}, \\
S_{klst}^{(1)} &= S_{kl}^{(1)}S_{st}^{(0)} + S_{kl}^{(0)}S_{st}^{(1)}.
\end{aligned} \tag{8}
$$

To obtain accurate TSh frequency with two-dimensional equations from power series approximations, correction factors must be inserted into stress expressions of the Mindlin plate equations (Wang et al. 2005, Du et al. 2012). In this paper, we did not derive a new set of these factors for nonlinear analysis but adopting the earlier factors from linear equations instead (Wang et al. 2005, Du et al. 2012). The simple reason is that the linear terms always dominate the vibration frequency and eventually the correction scheme (Wang et al. 2005, Du et al. 2012). Besides, the objective of nonlinear analysis in this paper is focused on the procedure of nonlinear analysis of frequency instability.

We now utilize the three-dimensional variational equation of motion (Wang et al., 2009b; Wu et al., 2012b)

$$\int_V \left[ \left( T_{ij} + T_{ik} u_{j,k} \right)_{,i} - \rho \ddot{u}_j \right] \delta u_j dV = 0 , \tag{9}$$

by substituting(1) into (9) and integrating through the thickness coordinate, then (9) becomes (Wu et al. 2012)

$$\int_A \sum_n \left[ \left( T_{ij}^{(n)} + \sum_{m=0}^{\infty} T_{ik}^{(m+n)} \bar{u}_{j,k}^{(m)} \right)_{,i} - nT_{2j}^{(n-1)} + F_j^{(n)} + \bar{F}_k^{(n)} - nT_{2k}^{(m+n-1)} \sum_{m=0}^{\infty} \bar{u}_{j,k}^{(m)} - \rho \sum_{m=0}^{\infty} B_{mn} \ddot{u}_j^{(m)} \right] \delta u_j^{(n)} dA = 0, \tag{10}$$

where

$$\begin{aligned} T_{ij}^{(n)} &= \int_{-b}^{b} T_{ij} x_2^n \mathrm{d}x_2 , \\ \bar{F}_k^{(n)} &= b^{m+n} T_{2k}\left(b\right) \sum_{m=0}^{\infty} \bar{u}_{j,k}^{(m)} - \left(-b\right)^{m+n} T_{2k}\left(-b\right) \sum_{m=0}^{\infty} \bar{u}_{j,k}^{(m)} , \\ F_j^{(n)} &= b^n T_{2j}\left(b\right) - \left(-b\right)^n T_{2j}\left(-b\right), \\ \bar{u}_{j,k}^{(m)} &= u_{j,k}^{(m)} + \left(m+1\right) \delta_{2k} u_j^{(m+1)} . \end{aligned} \tag{11}$$

With the divergence theorem, we have

$$\int_A \left( T_{ij}^{(n)} + \sum_{m=0}^{\infty} T_{ik}^{(m+n)} \bar{u}_{j,k}^{(m)} \right)_{,i} \delta u_j^{(n)} dA = \int_C \left( n_i T_{ij}^{(n)} + \sum_{m=0}^{\infty} n_i T_{ik}^{(m+n)} \bar{u}_{j,k}^{(m)} \right) \delta u_j^{(n)} \mathrm{d}S - \int_A \left( T_{ij}^{(n)} + \sum_{m=0}^{\infty} T_{ik}^{(m+n)} \bar{u}_{j,k}^{(m)} \right) \delta u_{j,i}^{(n)} dA. \tag{12}$$

With the definition of two-dimensional strains in (5), we can obtain

$$\sum_n T_{ij}^{(n)}\delta S_{ij}^{(n)} = \sum_n\left( T_{ij}^{(n)}\delta u_{j,i}^{(n)} + nT_{2j}^{(n-1)}\delta u_j^{(n)} + \sum_{m=0}^{\infty} T_{ik}^{(m+n)}\bar{u}_{j,k}^{(m)}\delta u_{j,i}^{(n)} + nT_{2k}^{(m+n-1)}\sum_{m=0}^{\infty}\bar{u}_{j,k}^{(m)}\delta u_j^{(n)}\right). \quad (13)$$

Through applying (12) and (13), (10) can be rewritten to

$$\int_A \sum_n\left( T_{ij}^{(n)}\delta S_{ij}^{(n)} + \rho\sum_{m=0}^{\infty} B_{mn}\ddot{u}_j^{(m)}\delta u_j^{(n)}\right)\mathrm{d}A = \int_C \sum_n\left( n_i T_{ij}^{(n)} + \sum_{m=0}^{\infty} n_i T_{ik}^{(m+n)}\bar{u}_{j,k}^{(m)}\right)\mathrm{d}S + \int_A \sum_n\left( F_j^{(n)} + \bar{F}_k^{(n)}\right)\mathrm{d}A, \quad (14)$$

The first and second terms of the right hand side in (14) represents the effect of external forces on edges and faces of plate, respectively. For free vibrations with traction-free edges and faces, these two terms will vanish, and (14) becomes

$$\int_A \sum_n\left( T_{ij}^{(n)}\delta S_{ij}^{(n)} + \rho\sum_{m=0}^{\infty} B_{mn}\ddot{u}_j^{(m)}\delta u_j^{(n)}\right)\mathrm{d}A = 0. \quad (15)$$

We shall use this equation for the formulation of finite element implementation of the nonlinear Mindlin plate equations.

## 3. NONLINEAR FINITE ELEMENT FORMULATION

As observed from the linear analysis of high frequency vibrations of plates, the TSh, flexural, and face-shear vibrations are the most strongly coupled modes in AT-cut rectangular quartz crystal plates (Wang and Zhao, 2005). As a result, corresponding displacement variables $u_2^{(0)}$, $u_3^{(0)}$ and $u_1^{(1)}$ should be kept as essential variables in nonlinear analysis also. In addition, since vibrations of a quartz plate can be more complicated, and may involve more coupled modes beyond these three important ones, we, therefore, need consider all the zeroth- and first-order displacements, strains, and stresses for the first-order finite element implementation of the nonlinear Mindlin plate theory. Thus, the corresponding vectors of variables are defined as (Wang et al., 1999, 2000)

$$
\begin{aligned}
\mathbf{u} &= \left\{u_1^{(0)}, u_2^{(0)}, u_3^{(0)}, u_1^{(1)}, u_2^{(1)}, u_3^{(1)}\right\}^{\mathbf{T}}, \\
\mathbf{S} &= \left\{S_1^{(0)}, S_2^{(0)}, S_3^{(0)}, S_4^{(0)}, S_5^{(0)}, S_6^{(0)}, S_1^{(1)}, S_2^{(1)}, S_3^{(1)}, S_4^{(1)}, S_5^{(1)}, S_6^{(1)}\right\}^{\mathbf{T}}, \\
\mathbf{T} &= \left\{T_1^{(0)}, T_2^{(0)}, T_3^{(0)}, T_4^{(0)}, T_5^{(0)}, T_6^{(0)}, T_1^{(1)}, T_2^{(1)}, T_3^{(1)}, T_4^{(1)}, T_5^{(1)}, T_6^{(1)}\right\}^{\mathbf{T}}.
\end{aligned} \tag{16}
$$

A standard finite element procedure assumes

$$
\mathbf{u} = \mathbf{N}\mathbf{U}, \tag{17}
$$

where **N** and **U** are the shape function matrix and node displacement vector, respectively.

Based on (3) and (4), the strain vector will be

$$
\begin{aligned}
\mathbf{S} &= (\mathbf{B_L} + \frac{1}{2}\mathbf{B_{NL}})\mathbf{U}, \\
\boldsymbol{\delta}\mathbf{S} &= (\mathbf{B_L} + \bar{\mathbf{B}}_{\mathbf{NL}})\boldsymbol{\delta}\mathbf{U},
\end{aligned} \tag{18}
$$

where $\mathbf{B}_{\mathrm{L}}$, $\mathbf{B}_{\mathrm{NL}}$, and $\bar{\mathbf{B}}_{\mathbf{NL}}$ are the linear strain matrix, nonlinear strain matrix, nonlinear differential strain matrix, respectively. $\mathbf{B}_{\mathrm{L}}$ is identical with the strain matrix appeared in our earlier linear analysis (Wang et al., 1999). It should be noted that any nonzero element in both $\mathbf{B}_{\mathrm{NL}}$ and $\bar{\mathbf{B}}_{\mathbf{NL}}$ (see Appendix) contains a displacement variable, which can represent the kinematic nonlinearity.

From (7), we can formulate the constitutive relations into matrix form

$$
\mathbf{T} = (\mathbf{D_L} + \mathbf{D_{NL}})\mathbf{S}, \tag{19}
$$

where $\mathbf{D}_{\mathrm{L}}$ and $\mathbf{D}_{\mathrm{NL}}$ are matrices of the linear and nonlinear parts of constitutive relations. $\mathbf{D}_{\mathrm{NL}}$ in which every nonzero element possesses a strain variable, can be divided into $\mathbf{D}_{\mathbf{NL}}^{\mathbf{S_L}}$ for the linear part and $\mathbf{D}_{\mathbf{NL}}^{\mathbf{S_{NL}}}$ for the nonlinear part of strain. That also means nonzero elements in $\mathbf{D}_{\mathbf{NL}}^{\mathbf{S_L}}$ will be with displacements in their first-orders, and nonzero elements in $\mathbf{D}_{\mathbf{NL}}^{\mathbf{S_{NL}}}$ will be with quadratic displacements on the other side. Again, expressions of $\mathbf{D}_{\mathbf{NL}}^{\mathbf{S_L}}$ and $\mathbf{D}_{\mathbf{NL}}^{\mathbf{S_{NL}}}$ are given in the Appendix. With this substitution, (19) becomes

$$\mathbf{T} = (\mathbf{D_L} + \mathbf{D_{NL}^{S_L}} + \mathbf{D_{NL}^{S_{NL}}})\mathbf{S}. \tag{20}$$

With definitions above, the weak form (15) can be written as

$$\int_A \boldsymbol{\delta}\mathbf{S}^\mathbf{T}\mathbf{T}\mathrm{d}A + \int_A \boldsymbol{\delta}\mathbf{u}^\mathbf{T}\mathbf{m}\ddot{\mathbf{u}}\mathrm{d}A = \mathbf{0}, \tag{21}$$

where **m** is the mass matrix, which is identical with our earlier linear analysis (Wang et al., 1999). Substituting (18) and (20) into (21), finally we have

$$\begin{aligned}(\mathbf{K_L} + \mathbf{K_{NL1}} + \mathbf{K_{NL2}} + \mathbf{K_{NL3}} + \mathbf{K_{NL4}} + \mathbf{K_{NL5}} + \mathbf{K_{NL6}} + \mathbf{K_{NL7}} + \mathbf{K_{NL8}} + \mathbf{K_{NL9}} + \mathbf{K_{NL10}} + \mathbf{K_{NL11}})\mathbf{U} \\ + \mathbf{M}\ddot{\mathbf{U}} = \mathbf{0},\end{aligned} \tag{22}$$

where

$$\begin{aligned}
&\mathbf{K_L} = \int_A \mathbf{B_L^T D_L B_L}\mathrm{d}A, \mathbf{K_{NL1}} = \int_A \mathbf{\bar{B}_{NL}^T D_L B_L}\mathrm{d}A, \mathbf{K_{NL2}} = \frac{1}{2}\int_A \mathbf{B_L^T D_L B_{NL}}\mathrm{d}A, \\
&\mathbf{K_{NL3}} = \frac{1}{2}\int_A \mathbf{\bar{B}_{NL}^T D_L B_{NL}}\mathrm{d}A, \mathbf{K_{NL4}} = \int_A \mathbf{B_L^T D_{NL}^{S_L} B_L}\mathrm{d}A, \mathbf{K_{NL5}} = \int_A \mathbf{\bar{B}_{NL}^T D_{NL}^{S_L} B_L}\mathrm{d}A, \\
&\mathbf{K_{NL6}} = \frac{1}{2}\int_A \mathbf{B_L^T D_{NL}^{S_L} B_{NL}}\mathrm{d}A, \mathbf{K_{NL7}} = \frac{1}{2}\int_A \mathbf{\bar{B}_{NL}^T D_{NL}^{S_L} B_{NL}}\mathrm{d}A, \mathbf{K_{NL8}} = \int_A \mathbf{B_L^T D_{NL}^{S_{NL}} B_L}\mathrm{d}A, \\
&\mathbf{K_{NL9}} = \int_A \mathbf{\bar{B}_{NL}^T D_{NL}^{S_{NL}} B_L}\mathrm{d}A, \mathbf{K_{NL10}} = \frac{1}{2}\int_A \mathbf{B_L^T D_{NL}^{S_{NL}} B_{NL}}\mathrm{d}A, \mathbf{K_{NL11}} = \frac{1}{2}\int_A \mathbf{\bar{B}_{NL}^T D_{NL}^{S_{NL}} B_{NL}}\mathrm{d}A, \\
&\mathbf{M} = \int_A \mathbf{N^T m N}\mathrm{d}A.
\end{aligned} \tag{23}$$

The matrix $\mathbf{K}_\mathrm{L}$ is the linear stiffness matrix and is the same as in our earlier linear analysis (Wang et al., 1999). Matrices $\mathbf{K}_\mathrm{NL1}$ to $\mathbf{K}_\mathrm{NL11}$ are nonlinear stiffness matrices due to nonlinear equations of strains and stresses.

In the first-order harmonic time domain approximation (Han and Petyt, 1997a, 1997b), displacements are assumed as

$$\mathbf{U} = \cos(\omega t)\mathbf{U_{max}}, \tag{24}$$

where $\omega$, $t$, and $\mathbf{U}_\mathrm{max}$ are vibration frequency, time, and the nonlinear vibration mode shape, respectively. Considering the orders of displacements in $\mathbf{B}_\mathrm{NL}$, $\mathbf{\bar{B}_{NL}}$, $\mathbf{D_{NL}^{S_L}}$ and $\mathbf{D_{NL}^{S_{NL}}}$, (22) becomes

$$[\mathbf{K}_{\mathbf{L}}\cos(\omega t)+\mathbf{K}_{\mathbf{NL1}}\cos^2(\omega t)+\mathbf{K}_{\mathbf{NL2}}\cos^2(\omega t)+\mathbf{K}_{\mathbf{NL3}}\cos^3(\omega t)+\mathbf{K}_{\mathbf{NL4}}\cos^2(\omega t)$$
$$+\mathbf{K}_{\mathbf{NL5}}\cos^3(\omega t)+\mathbf{K}_{\mathbf{NL6}}\cos^3(\omega t)+\mathbf{K}_{\mathbf{NL7}}\cos^4(\omega t)+\mathbf{K}_{\mathbf{NL8}}\cos^3(\omega t)+\mathbf{K}_{\mathbf{NL9}}\cos^4(\omega t) \quad (25)$$
$$+\mathbf{K}_{\mathbf{NL10}}\cos^4(\omega t)+\mathbf{K}_{\mathbf{NL11}}\cos^5(\omega t)]\mathbf{U}_{\mathbf{max}} - \omega^2\mathbf{M}\mathbf{U}_{\mathbf{max}}\cos(\omega t)=\mathbf{0}.$$

Here, we neglected effects from some higher-order harmonics, such as $\cos(2\omega t)$, $\cos(3\omega t)$, $\cos(4\omega t)$, and $\cos(5\omega t)$, and obtained the final matrix amplitude equation as

$$\left(\mathbf{K}_{\mathbf{L}}+\frac{3}{4}\mathbf{K}_{\mathbf{NL3}}+\frac{3}{4}\mathbf{K}_{\mathbf{NL5}}+\frac{3}{4}\mathbf{K}_{\mathbf{NL6}}+\frac{3}{4}\mathbf{K}_{\mathbf{NL8}}-\frac{5}{16}\mathbf{K}_{\mathbf{NL11}}\right)\mathbf{U}_{\max}-\omega^2\mathbf{M}\mathbf{U}_{\max}=\mathbf{0}. \quad (26)$$

Apparently, (26) is a generalized nonlinear eigensystem for $\omega$ and $\mathbf{U}_{\max}$, in which $\mathbf{K}_{\mathbf{NL3}}$, $\mathbf{K}_{\mathbf{NL5}}$, $\mathbf{K}_{\mathbf{NL6}}$, $\mathbf{K}_{\mathbf{NL8}}$ and $\mathbf{K}_{\mathbf{NL11}}$ are functions of $\mathbf{U}_{\max}$. Although $\omega$ and $\mathbf{U}_{\max}$ are just one possible set of solutions at given time of vibrations, it can provide a feasible way to study the steady vibration behavior of plates.

This is a standard nonlinear eigenvalue system from nonlinear vibrations of plates, but it is also commonly encountered in structural vibrations. Problems like this have been studied before, and we shall need to use effective and sophisticated methods from earlier studies for solutions of nonlinear vibrations with the objective of frequency stability analysis.

## 4. SOLUTION ALGORITHMS OF THE NONLINEAR EIGENSYSTEM

To solve the nonlinear eigensystem, or the so-called matrix amplitude equation in (26), we need to validate the accuracy of numerical solutions first. Currently, a nonlinear eigensystem like what we have could not be solved directly for one or several pairs of eigenvalues and eigenvectors. As a result, incremental iterations, such as Newton–Raphson procedure and arc-length or continuation method, are utilized frequently in the weakly or strongly coupled complex nonlinear analysis. This paper, however, employs a new 2-D method to study the frequency stability of resonators which focuses on frequency-amplitude relations of TSh vibrations, which are different from the nonlinear internal resonance and complicated non-monotonic backbone curves. Besides, we only consider one harmonic in time domain. The straightly direct iterations, consequently, would be enough for our nonlinear analysis. With this approach, only one eigenvalue and its corresponding eigenvector can be valid for one nonlinear eigensystem, although a set of eigenvalues and eigenvectors is solved by a linear procedure. The reason is that an eigenvector of a certain mode must be assembled into nonlinear stiffness matrices to form a new linearized eigensystem, which makes other eigenvalues and eigenvectors are not synchronized with their own modes in this linear eigensystem.

The iterative algorithms used in this paper can be described as

**STEP 1**. Solve the linear eigensystem (amplitude **U**= 0, set all elements in nonlinear stiffness matrices to be zero).

**STEP 2**.Search for the TSh frequency, and scale the corresponding eigenvector to a required amplitude.

**STEP 3**. Check the convergence criterion on TSh frequency ($\frac{F_{\text{present}} - F_{\text{last}}}{F_{\text{last}}} \leq 10^{-7}$). If it is satisfied, then move to STEP 6.

**STEP 4**. Formulate the nonlinear stiffness matrices with the eigenvector from STEP 2, and form a new linearized eigensystem.

**STEP 5**. Solve the new eigensystem by a linear solver, and obtain nonlinear frequencies and mode shapes. Then return to STEP 2.

**STEP 6**. Output results and terminate the program.

The linear eigensolver here is a new distributed parallel version especially for high-frequency analysis. It is a combination of four parallel high performance computational software components which includes `PARPACK`, `SuperLU_DIST`, `ParMETIS` and `Aztec` (Wang et al., 2011a).

## 5. RESULTS AND DISCUSSIONS

With a finite element program based on the formulation and implementation outlined above, we are ready to examine the results with a rectangular AT-cut quartz crystal plate from a real resonator model that has been analyzed with our linear program before (Wang et al., 2011a). The dimensions of plate are $2a$ = 1400μm, $2b$ = 41.1589μm, $2c$ = 1030μm with the plated quartz crystal blank having a center TSh frequency of 39.224030 MHz. Free vibration analysis with nonlinear finite element method usually gives two important results: backbone curves and nonlinear mode shapes at certain time ($\mathbf{U}_{max}$). In the beginning of finite element analysis, a convergence check should be made with different mesh schemes and initial DOFs are set around 180, 000. For this particular example, numerical results are presented in Table 1, in which "amplitude" is the numerical value of $u_1^{(1)}$ in the center of a plate and therefore can represent the TSh vibration amplitude. As already known, the largest TSh deformation usually happens in the middle of a rectangular plate. The largest TSh displacements are on the upper or bottom surface of plate. Their largest values will be around Amplitude$\times b$. In resonators, the maximum TSh displacements are usually very small, or in nanometer range. The "amplitude" in this analysis is 0.0050, which means actual deformation amplitude about $10^{-7}$ m. This represents an unusually large shear deformation in a quartz crystal plate. If the discrepancy of frequency-shifts is less than 0.1 ppm from analyses of two different meshes, then we assume the results are convergent, also means a 4Hz frequency difference in a 40MHz resonator. From Table 1, the acceptable results will need about 400, 000 DOFs for the nonlinear analysis.

Table 1 Convergence check for the nonlinear finite element analysis

| | Case 1 | Case 2 | Case 3 | Case 4 |
|---|---|---|---|---|
| Nodes in Length ×Nodes in Width | 227×137 | 285×165 | 341×193 | 413×273 |
| Number of DOFs | 186594 | 282150 | 394878 | 676494 |
| Frequency-Shift (ppm) Amplitude=0.0025 | 3.5732 | 3.3876 | 3.3316 | 3.2950 |
| Frequency-Shift(ppm) Amplitude=0.0050 | 14.2921 | 13.6388 | 13.4864 | 13.4129 |

In Fig. 2, the plot shows the frequency-amplitudes relations of TSh vibrations of a quartz crystal plate in which the amplitudes are the actual TSh displacements of the center of plate. As expected, the frequency increased as amplitudes became larger. It is very similar with the results from homotopy analysis (Wu et al., 2012b), which is based on the same nonlinear Mindlin plate theory but much simplified equations with the TSh vibrations considered only. This backbone curve is also similar to experimental and 3-D simulated DLD curves of resonators (Patel, 2008; Patel et al., 2009). Additionally, we compare two nonlinear TSh mode shapes with the linear mode shape in Fig. 3, in which the nonlinear TSh mode shapes with different amplitudes are scaled to the same magnitude at the center of plate. It is hard to observe the effect of nonlinearities from differences in displacements, even though the amplitudes are set to be very large and unrealistic in resonators. We also plotted $u_1^{(1)}$ values on a plane in Fig. 4 to illustrate the distribution of TSh vibrations in an AT-cut quartz crystal plate. It is shown that TSh vibrations are stronger in the center of plate than in two ends, seemingly identical with analyses from ANSYS (Wang et al., 2009a).

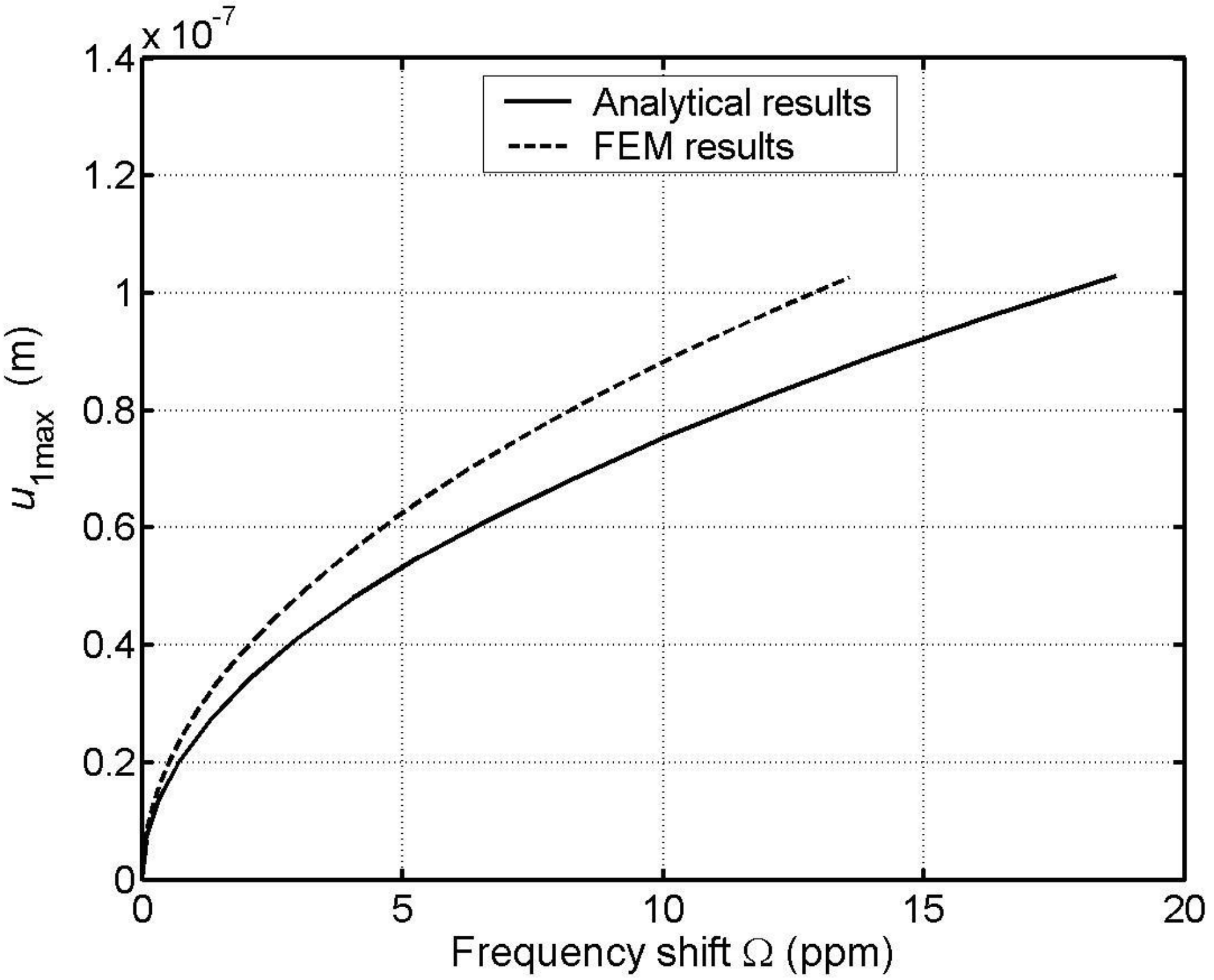

Fig. 2 Frequency-shifts at different amplitudes in comparison with nonlinear finite element analysis and analytical solution

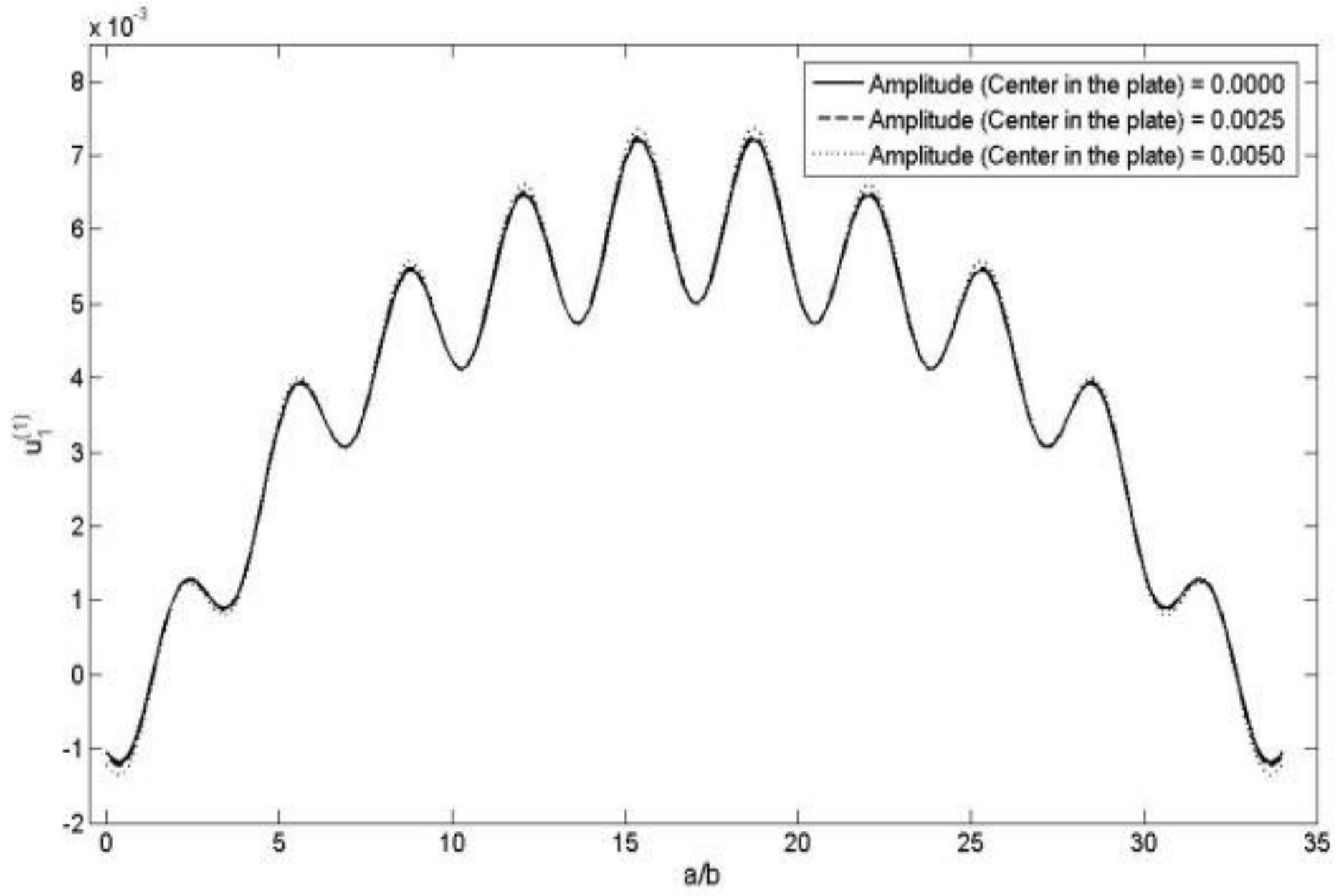

Fig. 3 Distributions of TSh vibrations in the central line of a plate with different amplitudes

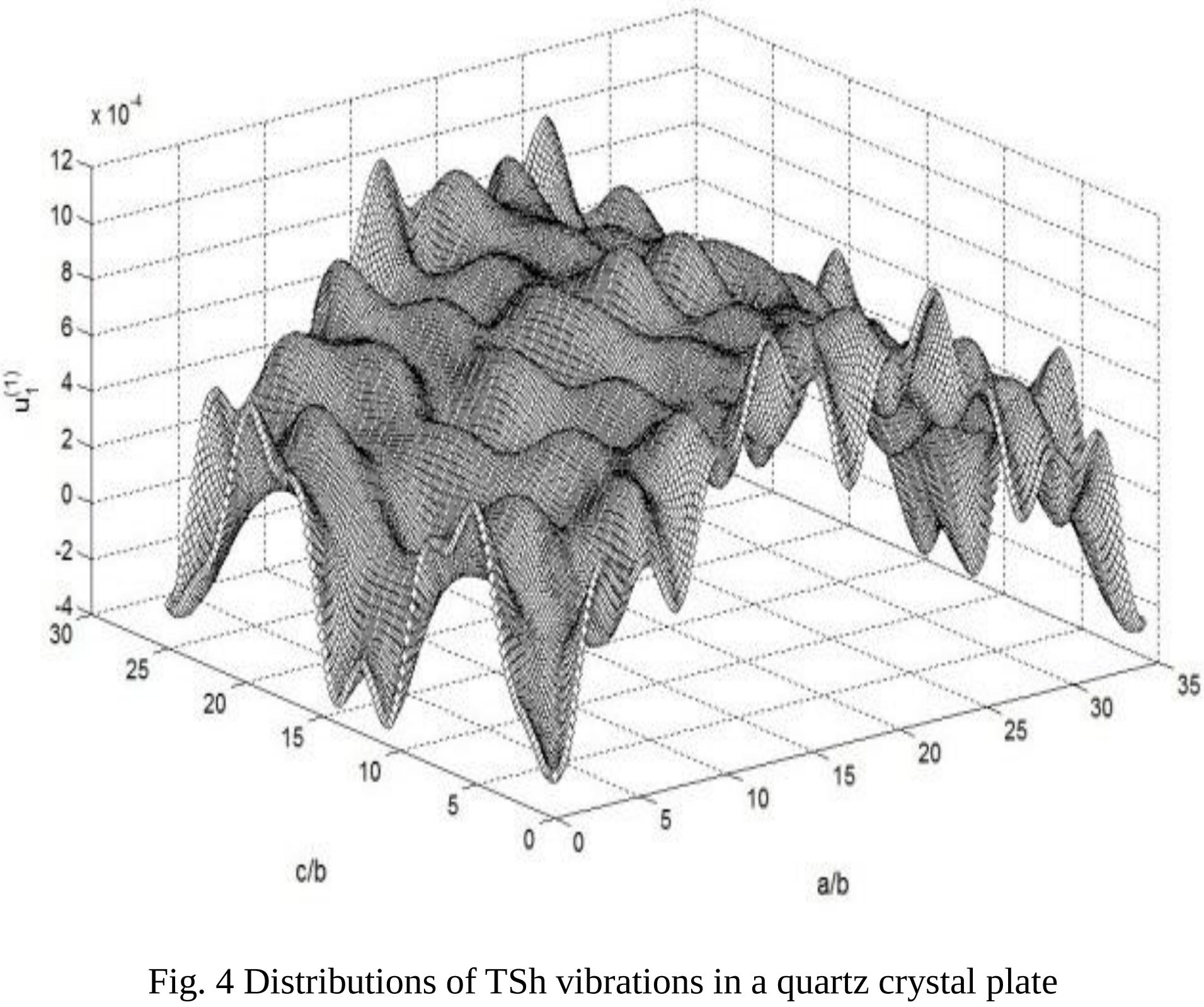

Fig. 4 Distributions of TSh vibrations in a quartz crystal plate

## 6. CONCLUSIONS

We presented a two-dimensional nonlinear finite element method to study TSh vibrations of quartz crystal plates. The Mindlin plate theory with the consideration of both kinematic and material nonlinearities has been established in our earlier papers and solved analytically under simplifications for some important results. As a systematic scheme, this nonlinear finite element implementation calculated the backbone curve, which is important in the analysis of frequency stability of quartz crystal resonators for improving certain critical properties such as the drive level dependency and thermal stability. The solutions from both finite element computation and approximate equations are consistent. A comparison between mode shapes of different amplitudes is made to examine the nonlinear effect in displacements. The amplitudes of TSh vibrations of resonators are extremely small as compared to thicknesses of quartz plates, and therefore TSh mode shapes would not change much by the inclusion of nonlinearities. Due to one harmonic time domain term approximation in our formulation, nonlinearities are included to make the plate stiffer in fact, thus increasing the frequency as well as adjusting the curvature of mode shapes at the same time. The internal resonance of complex nonlinear couplings cannot be analyzed in our programs, since it will need to consider more harmonics and apply incremental iterations. This nonlinear Mindlin plate theory based finite element method, as an expansion of the linear one, would be a basic tool in the design and analysis of quartz crystal resonators with enhanced capabilities for the examination of nonlinear properties.

**Acknowledgment**

The computing resources are provided by the Ningbo University-Dawning High Performance Computing Laboratory. This research is supported by the National Natural Science Foundation of China (Grant Nos.10772087, 10932004, and 11372145).

# APPENDIX

$$
\mathbf{B}_{\mathrm{NL}}=\begin{bmatrix}
\sum_{j=1}^{NEN}U_{(j)1}^{(0)}N_{i,1}N_{j,1} & \sum_{j=1}^{NEN}U_{(j)2}^{(0)}N_{i,1}N_{j,1} & \sum_{j=1}^{NEN}U_{(j)3}^{(0)}N_{i,1}N_{j,1} & 0 & 0 & 0 \\
0 & 0 & 0 & \sum_{j=1}^{NEN}U_{(j)1}^{(1)}N_{i}N_{j} & \sum_{j=1}^{NEN}U_{(j)2}^{(1)}N_{i}N_{j} & \sum_{j=1}^{NEN}U_{(j)3}^{(1)}N_{i}N_{j} \\
\sum_{j=1}^{NEN}U_{(j)1}^{(0)}N_{i,3}N_{j,3} & \sum_{j=1}^{NEN}U_{(j)2}^{(0)}N_{i,3}N_{j,3} & \sum_{j=1}^{NEN}U_{(j)3}^{(0)}N_{i,3}N_{j,3} & 0 & 0 & 0 \\
0 & 0 & 0 & 2\sum_{j=1}^{NEN}U_{(j)1}^{(0)}N_{i}N_{j,3} & 2\sum_{j=1}^{NEN}U_{(j)2}^{(0)}N_{i}N_{j,3} & 2\sum_{j=1}^{NEN}U_{(j)3}^{(0)}N_{i}N_{j,3} \\
2\sum_{j=1}^{NEN}U_{(j)1}^{(0)}N_{i,1}N_{j,3} & 2\sum_{j=1}^{NEN}U_{(j)2}^{(0)}N_{i,1}N_{j,3} & 2\sum_{j=1}^{NEN}U_{(j)3}^{(0)}N_{i,1}N_{j,3} & 0 & 0 & 0 \\
0 & 0 & 0 & 2\sum_{j=1}^{NEN}U_{(j)1}^{(0)}N_{i}N_{j,1} & 2\sum_{j=1}^{NEN}U_{(j)2}^{(0)}N_{i}N_{j,1} & 2\sum_{j=1}^{NEN}U_{(j)3}^{(0)}N_{i}N_{j,1} \\
0 & 0 & 0 & 2\sum_{j=1}^{NEN}U_{(j)1}^{(0)}N_{i,1}N_{j,1} & 2\sum_{j=1}^{NEN}U_{(j)2}^{(0)}N_{i,1}N_{j,1} & 2\sum_{j=1}^{NEN}U_{(j)3}^{(0)}N_{i,1}N_{j,1} \\
0 & 0 & 0 & 0 & 0 & 0 \\
0 & 0 & 0 & 2\sum_{j=1}^{NEN}U_{(j)1}^{(0)}N_{i,3}N_{j,3} & 2\sum_{j=1}^{NEN}U_{(j)2}^{(0)}N_{i,3}N_{j,3} & 2\sum_{j=1}^{NEN}U_{(j)3}^{(0)}N_{i,3}N_{j,3} \\
0 & 0 & 0 & 2\sum_{j=1}^{NEN}U_{(j)1}^{(1)}N_{i,3}N_{j} & 2\sum_{j=1}^{NEN}U_{(j)2}^{(1)}N_{i,3}N_{j} & 2\sum_{j=1}^{NEN}U_{(j)3}^{(1)}N_{i,3}N_{j} \\
0 & 0 & 0 & 2\sum_{j=1}^{NEN}U_{(j)1}^{(1)}\left(N_{i,1}N_{j,3}+N_{i,3}N_{j,1}\right) & 2\sum_{j=1}^{NEN}U_{(j)2}^{(1)}\left(N_{i,1}N_{j,3}+N_{i,3}N_{j,1}\right) & 2\sum_{j=1}^{NEN}U_{(j)3}^{(1)}\left(N_{i,1}N_{j,3}+N_{i,3}N_{j,1}\right) \\
0 & 0 & 0 & 2\sum_{j=1}^{NEN}U_{(j)1}^{(1)}N_{i,1}N_{j} & 2\sum_{j=1}^{NEN}U_{(j)2}^{(1)}N_{i,1}N_{j} & 2\sum_{j=1}^{NEN}U_{(j)3}^{(1)}N_{i,1}N_{j}
\end{bmatrix}_{12\times 6}
$$

$$
\bar{\mathbf{B}}_{\mathrm{NL}}=\begin{bmatrix}
\sum_{j=1}^{NEN}U_{(j)1}^{(0)}N_{i,1}N_{j,1} & \sum_{j=1}^{NEN}U_{(j)2}^{(0)}N_{i,1}N_{j,1} & \sum_{j=1}^{NEN}U_{(j)3}^{(0)}N_{i,1}N_{j,1} & 0 & 0 & 0 \\
0 & 0 & 0 & \sum_{j=1}^{NEN}U_{(j)1}^{(1)}N_{i}N_{j} & \sum_{j=1}^{NEN}U_{(j)2}^{(1)}N_{i}N_{j} & \sum_{j=1}^{NEN}U_{(j)3}^{(1)}N_{i}N_{j} \\
\sum_{j=1}^{NEN}U_{(j)1}^{(0)}N_{i,3}N_{j,3} & \sum_{j=1}^{NEN}U_{(j)2}^{(0)}N_{i,3}N_{j,3} & \sum_{j=1}^{NEN}U_{(j)3}^{(0)}N_{i,3}N_{j,3} & 0 & 0 & 0 \\
\sum_{j=1}^{NEN}U_{(j)1}^{(1)}N_{i,3}N_{j} & \sum_{j=1}^{NEN}U_{(j)2}^{(1)}N_{i,3}N_{j} & \sum_{j=1}^{NEN}U_{(j)3}^{(1)}N_{i,3}N_{j} & \sum_{j=1}^{NEN}U_{(j)1}^{(0)}N_{i}N_{j,3} & \sum_{j=1}^{NEN}U_{(j)2}^{(0)}N_{i}N_{j,3} & \sum_{j=1}^{NEN}U_{(j)3}^{(0)}N_{i}N_{j,3} \\
2\sum_{j=1}^{NEN}U_{(j)1}^{(0)}N_{i,1}N_{j,3} & 2\sum_{j=1}^{NEN}U_{(j)2}^{(0)}N_{i,1}N_{j,3} & 2\sum_{j=1}^{NEN}U_{(j)3}^{(0)}N_{i,1}N_{j,3} & 0 & 0 & 0 \\
\sum_{j=1}^{NEN}U_{(j)1}^{(1)}N_{i,1}N_{j} & \sum_{j=1}^{NEN}U_{(j)2}^{(1)}N_{i,1}N_{j} & \sum_{j=1}^{NEN}U_{(j)3}^{(1)}N_{i,1}N_{j} & \sum_{j=1}^{NEN}U_{(j)1}^{(0)}N_{i}N_{j,1} & \sum_{j=1}^{NEN}U_{(j)2}^{(0)}N_{i}N_{j,1} & \sum_{j=1}^{NEN}U_{(j)3}^{(0)}N_{i}N_{j,1} \\
\sum_{j=1}^{NEN}U_{(j)1}^{(1)}N_{i,1}N_{j,1} & \sum_{j=1}^{NEN}U_{(j)2}^{(1)}N_{i,1}N_{j,1} & \sum_{j=1}^{NEN}U_{(j)3}^{(1)}N_{i,1}N_{j,1} & \sum_{j=1}^{NEN}U_{(j)1}^{(0)}N_{i,1}N_{j,1} & \sum_{j=1}^{NEN}U_{(j)2}^{(0)}N_{i,1}N_{j,1} & \sum_{j=1}^{NEN}U_{(j)3}^{(0)}N_{i,1}N_{j,1} \\
0 & 0 & 0 & 0 & 0 & 0 \\
\sum_{j=1}^{NEN}U_{(j)1}^{(1)}N_{i,3}N_{j,3} & \sum_{j=1}^{NEN}U_{(j)2}^{(1)}N_{i,3}N_{j,3} & \sum_{j=1}^{NEN}U_{(j)3}^{(1)}N_{i,3}N_{j,3} & \sum_{j=1}^{NEN}U_{(j)1}^{(0)}N_{i,3}N_{j,3} & \sum_{j=1}^{NEN}U_{(j)2}^{(0)}N_{i,3}N_{j,3} & \sum_{j=1}^{NEN}U_{(j)3}^{(0)}N_{i,3}N_{j,3} \\
0 & 0 & 0 & 2\sum_{j=1}^{NEN}U_{(j)1}^{(1)}N_{i,3}N_{j} & 2\sum_{j=1}^{NEN}U_{(j)2}^{(1)}N_{i,3}N_{j} & 2\sum_{j=1}^{NEN}U_{(j)3}^{(1)}N_{i,3}N_{j} \\
\sum_{j=1}^{NEN}U_{(j)1}^{(1)}\left(N_{i,3}N_{j,1}+N_{i,1}N_{j,3}\right) & \sum_{j=1}^{NEN}U_{(j)2}^{(1)}\left(N_{i,3}N_{j,1}+N_{i,1}N_{j,3}\right) & \sum_{j=1}^{NEN}U_{(j)3}^{(1)}\left(N_{i,3}N_{j,1}+N_{i,1}N_{j,3}\right) & \sum_{j=1}^{NEN}U_{(j)1}^{(0)}\left(N_{i,3}N_{j,1}+N_{i,1}N_{j,3}\right) & \sum_{j=1}^{NEN}U_{(j)2}^{(0)}\left(N_{i,3}N_{j,1}+N_{i,1}N_{j,3}\right) & \sum_{j=1}^{NEN}U_{(j)3}^{(0)}\left(N_{i,3}N_{j,1}+N_{i,1}N_{j,3}\right) \\
0 & 0 & 0 & 2\sum_{j=1}^{NEN}U_{(j)1}^{(1)}N_{i,1}N_{j} & 2\sum_{j=1}^{NEN}U_{(j)2}^{(1)}N_{i,1}N_{j} & 2\sum_{j=1}^{NEN}U_{(j)3}^{(1)}N_{i,1}N_{j}
\end{bmatrix}_{12\times 6}
$$

$$
\mathbf{D}_{\mathrm{NL}}^{\mathrm{S_L}}=\begin{bmatrix}
\sum_{n=1}^{6}bS_{Ln}^{(0)}c_{1n1} & \sum_{n=1}^{6}bS_{Ln}^{(0)}c_{1n2} & \sum_{n=1}^{6}bS_{Ln}^{(0)}c_{1n3} & \sum_{n=1}^{6}bS_{Ln}^{(0)}c_{1n4} & \sum_{n=1}^{6}bS_{Ln}^{(0)}c_{1n5} & \sum_{n=1}^{6}bS_{Ln}^{(0)}c_{1n6} & 0 & 0 & 0 & 0 & 0 & 0 \\
\sum_{n=1}^{6}bS_{Ln}^{(0)}c_{2n1} & \sum_{n=1}^{6}bS_{Ln}^{(0)}c_{2n2} & \sum_{n=1}^{6}bS_{Ln}^{(0)}c_{2n3} & \sum_{n=1}^{6}bS_{Ln}^{(0)}c_{2n4} & \sum_{n=1}^{6}bS_{Ln}^{(0)}c_{2n5} & \sum_{n=1}^{6}bS_{Ln}^{(0)}c_{2n6} & 0 & 0 & 0 & 0 & 0 & 0 \\
\sum_{n=1}^{6}bS_{Ln}^{(0)}c_{3n1} & \sum_{n=1}^{6}bS_{Ln}^{(0)}c_{3n2} & \sum_{n=1}^{6}bS_{Ln}^{(0)}c_{3n3} & \sum_{n=1}^{6}bS_{Ln}^{(0)}c_{3n4} & \sum_{n=1}^{6}bS_{Ln}^{(0)}c_{3n5} & \sum_{n=1}^{6}bS_{Ln}^{(0)}c_{3n6} & 0 & 0 & 0 & 0 & 0 & 0 \\
\sum_{n=1}^{6}\frac{1}{2}bS_{Ln}^{(0)}c_{4n1} & \sum_{n=1}^{6}\frac{1}{2}bS_{Ln}^{(0)}c_{4n2} & \sum_{n=1}^{6}\frac{1}{2}bS_{Ln}^{(0)}c_{4n3} & \sum_{n=1}^{6}\frac{1}{2}bS_{Ln}^{(0)}c_{4n4} & \sum_{n=1}^{6}\frac{1}{2}bS_{Ln}^{(0)}c_{4n5} & \sum_{n=1}^{6}\frac{1}{2}bS_{Ln}^{(0)}c_{4n6} & 0 & 0 & 0 & 0 & 0 & 0 \\
\sum_{n=1}^{6}\frac{1}{2}bS_{Ln}^{(0)}c_{5n1} & \sum_{n=1}^{6}\frac{1}{2}bS_{Ln}^{(0)}c_{5n2} & \sum_{n=1}^{6}\frac{1}{2}bS_{Ln}^{(0)}c_{5n3} & \sum_{n=1}^{6}\frac{1}{2}bS_{Ln}^{(0)}c_{5n4} & \sum_{n=1}^{6}\frac{1}{2}bS_{Ln}^{(0)}c_{5n5} & \sum_{n=1}^{6}\frac{1}{2}bS_{Ln}^{(0)}c_{5n6} & 0 & 0 & 0 & 0 & 0 & 0 \\
\sum_{n=1}^{6}\frac{1}{2}bS_{Ln}^{(0)}c_{6n1} & \sum_{n=1}^{6}\frac{1}{2}bS_{Ln}^{(0)}c_{6n2} & \sum_{n=1}^{6}\frac{1}{2}bS_{Ln}^{(0)}c_{6n3} & \sum_{n=1}^{6}\frac{1}{2}bS_{Ln}^{(0)}c_{6n4} & \sum_{n=1}^{6}\frac{1}{2}bS_{Ln}^{(0)}c_{6n5} & \sum_{n=1}^{6}\frac{1}{2}bS_{Ln}^{(0)}c_{6n6} & 0 & 0 & 0 & 0 & 0 & 0 \\
\sum_{n=1}^{6}\frac{b^3}{3}S_{Ln}^{(1)}c_{1n1} & \sum_{n=1}^{6}\frac{b^3}{3}S_{Ln}^{(1)}c_{1n2} & \sum_{n=1}^{6}\frac{b^3}{3}S_{Ln}^{(1)}c_{1n3} & \sum_{n=1}^{6}\frac{b^3}{3}S_{Ln}^{(1)}c_{1n4} & \sum_{n=1}^{6}\frac{b^3}{3}S_{Ln}^{(1)}c_{1n5} & \sum_{n=1}^{6}\frac{b^3}{3}S_{Ln}^{(1)}c_{1n6} & \sum_{n=1}^{6}\frac{b^3}{3}S_{Ln}^{(0)}c_{1n1} & \sum_{n=1}^{6}\frac{b^3}{3}S_{Ln}^{(0)}c_{1n2} & \sum_{n=1}^{6}\frac{b^3}{3}S_{Ln}^{(0)}c_{1n3} & \sum_{n=1}^{6}\frac{b^3}{3}S_{Ln}^{(0)}c_{1n4} & \sum_{n=1}^{6}\frac{b^3}{3}S_{Ln}^{(0)}c_{1n5} & \sum_{n=1}^{6}\frac{b^3}{3}S_{Ln}^{(0)}c_{1n6} \\
\sum_{n=1}^{6}\frac{b^3}{3}S_{Ln}^{(1)}c_{2n1} & \sum_{n=1}^{6}\frac{b^3}{3}S_{Ln}^{(1)}c_{2n2} & \sum_{n=1}^{6}\frac{b^3}{3}S_{Ln}^{(1)}c_{2n3} & \sum_{n=1}^{6}\frac{b^3}{3}S_{Ln}^{(1)}c_{2n4} & \sum_{n=1}^{6}\frac{b^3}{3}S_{Ln}^{(1)}c_{2n5} & \sum_{n=1}^{6}\frac{b^3}{3}S_{Ln}^{(1)}c_{2n6} & \sum_{n=1}^{6}\frac{b^3}{3}S_{Ln}^{(0)}c_{2n1} & \sum_{n=1}^{6}\frac{b^3}{3}S_{Ln}^{(0)}c_{2n2} & \sum_{n=1}^{6}\frac{b^3}{3}S_{Ln}^{(0)}c_{2n3} & \sum_{n=1}^{6}\frac{b^3}{3}S_{Ln}^{(0)}c_{2n4} & \sum_{n=1}^{6}\frac{b^3}{3}S_{Ln}^{(0)}c_{2n5} & \sum_{n=1}^{6}\frac{b^3}{3}S_{Ln}^{(0)}c_{2n6} \\
\sum_{n=1}^{6}\frac{b^3}{3}S_{Ln}^{(1)}c_{3n1} & \sum_{n=1}^{6}\frac{b^3}{3}S_{Ln}^{(1)}c_{3n2} & \sum_{n=1}^{6}\frac{b^3}{3}S_{Ln}^{(1)}c_{3n3} & \sum_{n=1}^{6}\frac{b^3}{3}S_{Ln}^{(1)}c_{3n4} & \sum_{n=1}^{6}\frac{b^3}{3}S_{Ln}^{(1)}c_{3n5} & \sum_{n=1}^{6}\frac{b^3}{3}S_{Ln}^{(1)}c_{3n6} & \sum_{n=1}^{6}\frac{b^3}{3}S_{Ln}^{(0)}c_{3n1} & \sum_{n=1}^{6}\frac{b^3}{3}S_{Ln}^{(0)}c_{3n2} & \sum_{n=1}^{6}\frac{b^3}{3}S_{Ln}^{(0)}c_{3n3} & \sum_{n=1}^{6}\frac{b^3}{3}S_{Ln}^{(0)}c_{3n4} & \sum_{n=1}^{6}\frac{b^3}{3}S_{Ln}^{(0)}c_{3n5} & \sum_{n=1}^{6}\frac{b^3}{3}S_{Ln}^{(0)}c_{3n6} \\
\sum_{n=1}^{6}\frac{b^3}{6}S_{Ln}^{(1)}c_{4n1} & \sum_{n=1}^{6}\frac{b^3}{6}S_{Ln}^{(1)}c_{4n2} & \sum_{n=1}^{6}\frac{b^3}{6}S_{Ln}^{(1)}c_{4n3} & \sum_{n=1}^{6}\frac{b^3}{6}S_{Ln}^{(1)}c_{4n4} & \sum_{n=1}^{6}\frac{b^3}{6}S_{Ln}^{(1)}c_{4n5} & \sum_{n=1}^{6}\frac{b^3}{6}S_{Ln}^{(1)}c_{4n6} & \sum_{n=1}^{6}\frac{b^3}{6}S_{Ln}^{(0)}c_{4n1} & \sum_{n=1}^{6}\frac{b^3}{6}S_{Ln}^{(0)}c_{4n2} & \sum_{n=1}^{6}\frac{b^3}{6}S_{Ln}^{(0)}c_{4n3} & \sum_{n=1}^{6}\frac{b^3}{6}S_{Ln}^{(0)}c_{4n4} & \sum_{n=1}^{6}\frac{b^3}{6}S_{Ln}^{(0)}c_{4n5} & \sum_{n=1}^{6}\frac{b^3}{6}S_{Ln}^{(0)}c_{4n6} \\
\sum_{n=1}^{6}\frac{b^3}{6}S_{Ln}^{(1)}c_{5n1} & \sum_{n=1}^{6}\frac{b^3}{6}S_{Ln}^{(1)}c_{5n2} & \sum_{n=1}^{6}\frac{b^3}{6}S_{Ln}^{(1)}c_{5n3} & \sum_{n=1}^{6}\frac{b^3}{6}S_{Ln}^{(1)}c_{5n4} & \sum_{n=1}^{6}\frac{b^3}{6}S_{Ln}^{(1)}c_{5n5} & \sum_{n=1}^{6}\frac{b^3}{6}S_{Ln}^{(1)}c_{5n6} & \sum_{n=1}^{6}\frac{b^3}{6}S_{Ln}^{(0)}c_{5n1} & \sum_{n=1}^{6}\frac{b^3}{6}S_{Ln}^{(0)}c_{5n2} & \sum_{n=1}^{6}\frac{b^3}{6}S_{Ln}^{(0)}c_{5n3} & \sum_{n=1}^{6}\frac{b^3}{6}S_{Ln}^{(0)}c_{5n4} & \sum_{n=1}^{6}\frac{b^3}{6}S_{Ln}^{(0)}c_{5n5} & \sum_{n=1}^{6}\frac{b^3}{6}S_{Ln}^{(0)}c_{5n6} \\
\sum_{n=1}^{6}\frac{b^3}{6}S_{Ln}^{(1)}c_{6n1} & \sum_{n=1}^{6}\frac{b^3}{6}S_{Ln}^{(1)}c_{6n2} & \sum_{n=1}^{6}\frac{b^3}{6}S_{Ln}^{(1)}c_{6n3} & \sum_{n=1}^{6}\frac{b^3}{6}S_{Ln}^{(1)}c_{6n4} & \sum_{n=1}^{6}\frac{b^3}{6}S_{Ln}^{(1)}c_{6n5} & \sum_{n=1}^{6}\frac{b^3}{6}S_{Ln}^{(1)}c_{6n6} & \sum_{n=1}^{6}\frac{b^3}{6}S_{Ln}^{(0)}c_{6n1} & \sum_{n=1}^{6}\frac{b^3}{6}S_{Ln}^{(0)}c_{6n2} & \sum_{n=1}^{6}\frac{b^3}{6}S_{Ln}^{(0)}c_{6n3} & \sum_{n=1}^{6}\frac{b^3}{6}S_{Ln}^{(0)}c_{6n4} & \sum_{n=1}^{6}\frac{b^3}{6}S_{Ln}^{(0)}c_{6n5} & \sum_{n=1}^{6}\frac{b^3}{6}S_{Ln}^{(0)}c_{6n6}
\end{bmatrix}_{12\times 12}
$$

$$\mathbf{D}_{\mathbf{NL}}^{\mathbf{S_{NL}}}=\begin{bmatrix}
\sum_{n=1}^{6}bS_{NL}^{(0)}c_{1n1} & \sum_{n=1}^{6}bS_{NL}^{(0)}c_{1n2} & \sum_{n=1}^{6}bS_{NL}^{(0)}c_{1n3} & \sum_{n=1}^{6}bS_{NL}^{(0)}c_{1n4} & \sum_{n=1}^{6}bS_{NL}^{(0)}c_{1n5} & \sum_{n=1}^{6}bS_{NL}^{(0)}c_{1n6} & 0 & 0 & 0 & 0 & 0 & 0 \\
\sum_{n=1}^{6}bS_{NL}^{(0)}c_{2n1} & \sum_{n=1}^{6}bS_{NL}^{(0)}c_{2n2} & \sum_{n=1}^{6}bS_{NL}^{(0)}c_{2n3} & \sum_{n=1}^{6}bS_{NL}^{(0)}c_{2n4} & \sum_{n=1}^{6}bS_{NL}^{(0)}c_{2n5} & \sum_{n=1}^{6}bS_{NL}^{(0)}c_{2n6} & 0 & 0 & 0 & 0 & 0 & 0 \\
\sum_{n=1}^{6}bS_{NL}^{(0)}c_{3n1} & \sum_{n=1}^{6}bS_{NL}^{(0)}c_{3n2} & \sum_{n=1}^{6}bS_{NL}^{(0)}c_{3n3} & \sum_{n=1}^{6}bS_{NL}^{(0)}c_{3n4} & \sum_{n=1}^{6}bS_{NL}^{(0)}c_{3n5} & \sum_{n=1}^{6}bS_{NL}^{(0)}c_{3n6} & 0 & 0 & 0 & 0 & 0 & 0 \\
\sum_{n=1}^{6}\frac{1}{2}bS_{NL}^{(0)}c_{4n1} & \sum_{n=1}^{6}\frac{1}{2}bS_{NL}^{(0)}c_{4n2} & \sum_{n=1}^{6}\frac{1}{2}bS_{NL}^{(0)}c_{4n3} & \sum_{n=1}^{6}\frac{1}{2}bS_{NL}^{(0)}c_{4n4} & \sum_{n=1}^{6}\frac{1}{2}bS_{NL}^{(0)}c_{4n5} & \sum_{n=1}^{6}\frac{1}{2}bS_{NL}^{(0)}c_{4n6} & 0 & 0 & 0 & 0 & 0 & 0 \\
\sum_{n=1}^{6}\frac{1}{2}bS_{NL}^{(0)}c_{5n1} & \sum_{n=1}^{6}\frac{1}{2}bS_{NL}^{(0)}c_{5n2} & \sum_{n=1}^{6}\frac{1}{2}bS_{NL}^{(0)}c_{5n3} & \sum_{n=1}^{6}\frac{1}{2}bS_{NL}^{(0)}c_{5n4} & \sum_{n=1}^{6}\frac{1}{2}bS_{NL}^{(0)}c_{5n5} & \sum_{n=1}^{6}\frac{1}{2}bS_{NL}^{(0)}c_{5n6} & 0 & 0 & 0 & 0 & 0 & 0 \\
\sum_{n=1}^{6}\frac{1}{2}bS_{NL}^{(0)}c_{6n1} & \sum_{n=1}^{6}\frac{1}{2}bS_{NL}^{(0)}c_{6n2} & \sum_{n=1}^{6}\frac{1}{2}bS_{NL}^{(0)}c_{6n3} & \sum_{n=1}^{6}\frac{1}{2}bS_{NL}^{(0)}c_{6n4} & \sum_{n=1}^{6}\frac{1}{2}bS_{NL}^{(0)}c_{6n5} & \sum_{n=1}^{6}\frac{1}{2}bS_{NL}^{(0)}c_{6n6} & 0 & 0 & 0 & 0 & 0 & 0 \\
\sum_{n=1}^{6}\frac{b^3}{3}S_{NL}^{(1)}c_{1n1} & \sum_{n=1}^{6}\frac{b^3}{3}S_{NL}^{(1)}c_{1n2} & \sum_{n=1}^{6}\frac{b^3}{3}S_{NL}^{(1)}c_{1n3} & \sum_{n=1}^{6}\frac{b^3}{3}S_{NL}^{(1)}c_{1n4} & \sum_{n=1}^{6}\frac{b^3}{3}S_{NL}^{(1)}c_{1n5} & \sum_{n=1}^{6}\frac{b^3}{3}S_{NL}^{(1)}c_{1n6} & \sum_{n=1}^{6}\frac{b^3}{3}S_{NL}^{(0)}c_{1n1} & \sum_{n=1}^{6}\frac{b^3}{3}S_{NL}^{(0)}c_{1n2} & \sum_{n=1}^{6}\frac{b^3}{3}S_{NL}^{(0)}c_{1n3} & \sum_{n=1}^{6}\frac{b^3}{3}S_{NL}^{(0)}c_{1n4} & \sum_{n=1}^{6}\frac{b^3}{3}S_{NL}^{(0)}c_{1n5} & \sum_{n=1}^{6}\frac{b^3}{3}S_{NL}^{(0)}c_{1n6} \\
\sum_{n=1}^{6}\frac{b^3}{3}S_{NL}^{(1)}c_{2n1} & \sum_{n=1}^{6}\frac{b^3}{3}S_{NL}^{(1)}c_{2n2} & \sum_{n=1}^{6}\frac{b^3}{3}S_{NL}^{(1)}c_{2n3} & \sum_{n=1}^{6}\frac{b^3}{3}S_{NL}^{(1)}c_{2n4} & \sum_{n=1}^{6}\frac{b^3}{3}S_{NL}^{(1)}c_{2n5} & \sum_{n=1}^{6}\frac{b^3}{3}S_{NL}^{(1)}c_{2n6} & \sum_{n=1}^{6}\frac{b^3}{3}S_{NL}^{(0)}c_{2n1} & \sum_{n=1}^{6}\frac{b^3}{3}S_{NL}^{(0)}c_{2n2} & \sum_{n=1}^{6}\frac{b^3}{3}S_{NL}^{(0)}c_{2n3} & \sum_{n=1}^{6}\frac{b^3}{3}S_{NL}^{(0)}c_{2n4} & \sum_{n=1}^{6}\frac{b^3}{3}S_{NL}^{(0)}c_{2n5} & \sum_{n=1}^{6}\frac{b^3}{3}S_{NL}^{(0)}c_{2n6} \\
\sum_{n=1}^{6}\frac{b^3}{3}S_{NL}^{(1)}c_{3n1} & \sum_{n=1}^{6}\frac{b^3}{3}S_{NL}^{(1)}c_{3n2} & \sum_{n=1}^{6}\frac{b^3}{3}S_{NL}^{(1)}c_{3n3} & \sum_{n=1}^{6}\frac{b^3}{3}S_{NL}^{(1)}c_{3n4} & \sum_{n=1}^{6}\frac{b^3}{3}S_{NL}^{(1)}c_{3n5} & \sum_{n=1}^{6}\frac{b^3}{3}S_{NL}^{(1)}c_{3n6} & \sum_{n=1}^{6}\frac{b^3}{3}S_{NL}^{(0)}c_{3n1} & \sum_{n=1}^{6}\frac{b^3}{3}S_{NL}^{(0)}c_{3n2} & \sum_{n=1}^{6}\frac{b^3}{3}S_{NL}^{(0)}c_{3n3} & \sum_{n=1}^{6}\frac{b^3}{3}S_{NL}^{(0)}c_{3n4} & \sum_{n=1}^{6}\frac{b^3}{3}S_{NL}^{(0)}c_{3n5} & \sum_{n=1}^{6}\frac{b^3}{3}S_{NL}^{(0)}c_{3n6} \\
\sum_{n=1}^{6}\frac{b^3}{6}S_{NL}^{(1)}c_{4n1} & \sum_{n=1}^{6}\frac{b^3}{6}S_{NL}^{(1)}c_{4n2} & \sum_{n=1}^{6}\frac{b^3}{6}S_{NL}^{(1)}c_{4n3} & \sum_{n=1}^{6}\frac{b^3}{6}S_{NL}^{(1)}c_{4n4} & \sum_{n=1}^{6}\frac{b^3}{6}S_{NL}^{(1)}c_{4n5} & \sum_{n=1}^{6}\frac{b^3}{6}S_{NL}^{(1)}c_{4n6} & \sum_{n=1}^{6}\frac{b^3}{6}S_{NL}^{(0)}c_{4n1} & \sum_{n=1}^{6}\frac{b^3}{6}S_{NL}^{(0)}c_{4n2} & \sum_{n=1}^{6}\frac{b^3}{6}S_{NL}^{(0)}c_{4n3} & \sum_{n=1}^{6}\frac{b^3}{6}S_{NL}^{(0)}c_{4n4} & \sum_{n=1}^{6}\frac{b^3}{6}S_{NL}^{(0)}c_{4n5} & \sum_{n=1}^{6}\frac{b^3}{6}S_{NL}^{(0)}c_{4n6} \\
\sum_{n=1}^{6}\frac{b^3}{6}S_{NL}^{(1)}c_{5n1} & \sum_{n=1}^{6}\frac{b^3}{6}S_{NL}^{(1)}c_{5n2} & \sum_{n=1}^{6}\frac{b^3}{6}S_{NL}^{(1)}c_{5n3} & \sum_{n=1}^{6}\frac{b^3}{6}S_{NL}^{(1)}c_{5n4} & \sum_{n=1}^{6}\frac{b^3}{6}S_{NL}^{(1)}c_{5n5} & \sum_{n=1}^{6}\frac{b^3}{6}S_{NL}^{(1)}c_{5n6} & \sum_{n=1}^{6}\frac{b^3}{6}S_{NL}^{(0)}c_{5n1} & \sum_{n=1}^{6}\frac{b^3}{6}S_{NL}^{(0)}c_{5n2} & \sum_{n=1}^{6}\frac{b^3}{6}S_{NL}^{(0)}c_{5n3} & \sum_{n=1}^{6}\frac{b^3}{6}S_{NL}^{(0)}c_{5n4} & \sum_{n=1}^{6}\frac{b^3}{6}S_{NL}^{(0)}c_{5n5} & \sum_{n=1}^{6}\frac{b^3}{6}S_{NL}^{(0)}c_{5n6} \\
\sum_{n=1}^{6}\frac{b^3}{6}S_{NL}^{(1)}c_{6n1} & \sum_{n=1}^{6}\frac{b^3}{6}S_{NL}^{(1)}c_{6n2} & \sum_{n=1}^{6}\frac{b^3}{6}S_{NL}^{(1)}c_{6n3} & \sum_{n=1}^{6}\frac{b^3}{6}S_{NL}^{(1)}c_{6n4} & \sum_{n=1}^{6}\frac{b^3}{6}S_{NL}^{(1)}c_{6n5} & \sum_{n=1}^{6}\frac{b^3}{6}S_{NL}^{(1)}c_{6n6} & \sum_{n=1}^{6}\frac{b^3}{6}S_{NL}^{(0)}c_{6n1} & \sum_{n=1}^{6}\frac{b^3}{6}S_{NL}^{(0)}c_{6n2} & \sum_{n=1}^{6}\frac{b^3}{6}S_{NL}^{(0)}c_{6n3} & \sum_{n=1}^{6}\frac{b^3}{6}S_{NL}^{(0)}c_{6n4} & \sum_{n=1}^{6}\frac{b^3}{6}S_{NL}^{(0)}c_{6n5} & \sum_{n=1}^{6}\frac{b^3}{6}S_{NL}^{(0)}c_{6n6}
\end{bmatrix}_{12\times 12}.$$

## References

Bathe, K.-J., Bolourchi, S., 1980. A geometric and material nonlinear plate and shell element, Comput. Struct. 11, 23-48.

Chen, S. H., Cheung, Y. K., Xing, H. X., 2001. Nonlinear vibration of plane structures by finite element and incremental harmonic balance method. Nonlinear Dynam. 26, 87–104.

Cheung, Y. K., Lau, S. L., 1982. Incremental time-space finite strip method for nonlinear structural vibrations, Earthq. Eng. Struct. Dyn. 10, 239-253.

Du, J. K., Chen, G. J., Wang, W. J., Wu, R. X., Ma, T. F., Wang, J., 2012. Correction Factors of the Mindlin Plate Equations with the Consideration of Electrodes, IEEE Trans. Ultrason. Ferroelect. and Freq. Contr. 59, 2352-2358.

Han, W., Petyt, M., 1997a. Geometrically nonlinear vibration analysis of thin, rectangular plates using the hierarchical finite element method - 1: the fundamental mode of isoreopic plates, Comput. Struct. 63, 295-308.

Han, W., Petyt, M., 1997b. Geometrically nonlinear vibration analysis of thin, rectangular plates using the hierarchical finite element method - 2: its mode of laminated plates and higher modes of isotropic and laminated plates, Comput. Struct., 63, 309-318.

Lau, S. L., Cheung, Y. K., Wu, S. Y., 1983. Incremental harmonic balance method with multiple time scales for aperiodic vibration of nonlinear system, Trans. ASME, J. Appl. Mech. 50, 871-876.

Lau, S. L., Cheung,Y. K., Wu, S. Y., 1984. Non-linear vibration of thin elastic plates. Part 2: internal resonance by amplitude- incremental finite element. Trans. ASME, J. Appl. Mech. 51, 845-85

Lewandowski, R., 1994. Non-linear free vibrations of beams by the finite element and continuation methods, J. Sound Vib. 170, 577–593.

Lewandowski, R., 1997a. Computational formulation for periodic vibration of geometrically nonlinear structures-part 1: theoretical background. Int. J. Solids. Struct., 34, 1925-1947.

Lewandowski, R., 1997b. Computational formulation for periodic vibration of geometrically nonlinear structures-part 2: numerical strategy and examples. Int. J. Solids. Struct., 34, 1949-1964.

Leung, A. Y. T., Fung, T. C., 1989. Nonlinear steady state vibration of frames by finite element method, Int. J. Numer. Meth. Eng., 28, 1599-1618.

Mei, C., 1972. Nonlinear vibrations of beams by matrix displacement method. AIAA J.10, 355-357.

Mei, C., 1973. Finite Element displacement method for large amplitude free flexural vibrations of beam and plates. Comput. Struct. 3, 163-174.

Mindlin, R. D., 2006. An Introduction to the Mathematical Theory of Vibrations of Elastic Plates. J. S. Yang Ed., Singapore: World Scientific.

Patel, M. S., 2008. Nonlinear behavior in quartz crystal resonators and its stability. Ph. D. Dissertation, Dept. of Civil and Environmental Engineering, Rutgers University, Newark, NJ.

Patel, M. S., Yong, Y.-K., Tanaka, M., 2009. Drive level dependency in quartz resonators. Int. J. Solids

Struct. 46, 1856-1871.

Putcha, N. S., Reddy, J. N., 1986. Stability and natural vibration analysis of laminated plates by using a mixed element based on a refined plate theory, J. Sound Vib. 104, 285-300.

Rao, G. V., Raju, K. K., Rajut, I.S., 1976. Finite element formulation for the large amplitude free vibrations of beams and orthotropic circular plates, Comput. Struct. 6, 169-172.

Ribeiro, P., Petyt, M., 1999a. Nonlinear vibration of plates by the hierarchical finite element and continuation methods, Int. J. Mech. Sci. 41, 437-459.

Ribeiro, P., Petyt, M., 1999b. Multi-modal geometrical non-linear free vibration of fully clamped composite laminated plates, J. Sound Vib. 225, 127-152.

Ribeiro, P., Petyt, M., 1999c. Non-linear vibration of beams with internal resonance by the hierarchical finite-element method. J. Sound Vib. 224, 591-624.

Ribeiro, P., Petyt, M., 2000. Non-linear free vibration of isotropic plates with internal resonance. Int. J. NonLin. Mech. 35, 263-278.

Singha, M. K., Daripa, R., 2007. Nonlinear vibration of symmetrically laminated composite skewplates by finite element method, Int. J. Nonlinear Mech. 42, 1144-1152.

Wang, J., Yong, Y.-K., Imai, T., 1999. Finite element analysis of the piezoelectric vibrations of quartz plate resonators with higher-order plate theory. Int. J. Solids Struct. 36, 2303-2030.

Wang, J., Yu, J.-D., Yong, Y.-K., Imai, T., 2000. A new theory for electroded piezoelectric plates and its finite element application for the forced vibrations of quartz crystal resonators. Int. J. Solids Struct. 37, 5653-5673.

Wang, J., Yang, J. S., 2000. Higher-order theories of piezoelectric plates and applications. Appl. Mech. Rev. 5387–5399.

Wang, J., Lin ,J. B., Wan,Y. P., Zhong, Z., 2005. A two-dimensional analysis of surface acoustic waves in finite solids with considerations of electrodes, Int. J. Appl. Electrom. 22, 53-68.

Wang, J., Zhao, W. H., 2005. The determination of the optimal lengthof crystal blanks in quartz crystal resonators. IEEE Trans. Ultrason. Ferroelectr. Freq. Control. 52, 2023-2030.

Wang, J., Shi, J., Du, J. K., 2009a. The finite element analysis of thickness-shear vibrations of quartz crystal plates with ANSYS. Proceedings of the Joint Conference of the 2009 Symposium on Piezoelectricity, Acoustic Waves, and Device Applications and China Symposium on Frequency Control Technology.

Wang, J., Wu, R.X., Du, J.K., 2009b. The nonlinear thickness-shear vibrations of an infinite and isotropic elastic plate. Proceedings of the Joint Conference of the 2009 Symposium on Piezoelectricity, Acoustic Waves, and Device Applications and 2009 China Symposium on Frequency Control Technology.

Wang, J., Chen, Y. Y., Du, J. K., Wang, L. H., Ma, T. F., Huang, D. J., 2011a. A Mindlin Plate Theory Based Parallel 2D Finite Element Analysis of Quartz Crystal Resonators. Proceedings of the 2011 IEEE International Ultrasonics Symposium.

Wang, J., Wu, R. X., Du, J. K., Yan, W., Huang, D. J., 2011b. Analytical solutions of nonlinear vibrations of thickness-shear and flexural modes of quartz plates. Proceedings of the 2011 IEEE International Frequency Control Symposium. 131-134.

Wu, R. X., Wang, J., Du, J. K., 2011. The nonlinear thickness-shear vibrations of quartz crystal plates under a strong electric field. Proceedings of the 2011 IEEE International Ultrasonics Symposium.

Wu, R . X., Wang, J., Du, J. K., Hu, Y. T., Hu, H. P., 2012a. Solutions of nonlinear thickness-shear vibrations of an infinite isotropic plate with the homotopy analysis method, Numer. Algorithms. 59,

213–226.

Wu, R. X., Wang, J., Du, J. K., Huang, D. J., Yan, W., Hu, Y. T., 2012b. An analysis of nonlinear vibrations of coupled thickness-shear and flexural modes of quartz crystal plates with the homotopy analysis method, IEEE Trans. Ultrason. Ferroelect. and Freq. Contr. 59, 30-39.

Yang, J.S., 1999. Equations for the extension and flexure of electroelastic plates under strong electric fields, Int. J. Solids Struct. 36, 3171-3192.